*Orientational Order in the Nematic and Heliconical Nematic Liquid Crystals*

Gautam Singh,[a] Dena M. Agra-Kooijman,[a] Michael R. Fisch,[b] M. R. Vengatesan,[c]
Jang-Kun Song,[c] and Satyendra Kumar [a]

[a]Department of Physics and [b]College of Applied Engineering Sustainability and Technology
Kent State University, Kent, OH 44242

[c]School of Electronic and Electrical Engineering, Sungkyunkwan University
Suwon, Republic of Korea

Corresponding Author:
Satyendra Kumar
Department of Physics, Kent State University, Kent, OH 44242
Email: skumar@kent.edu
Phone: (330) 672-2566

**Abstract**

X-ray scattering and polarized microscopic studies of the structure and order parameters in the nematic (N) and heliconical, or the twist-bend nematic ($N_{tb}$), phase have been performed as a function of temperature. The nematic orientational order parameters $<P_2(\cos\theta)>$ and $<P_4(\cos\theta)>$ in the nematic phases of CB7CB and its mixtures with less than 20 wt% CB6CB reveal that they both increase with decreasing temperature in the N phase. Both order parameters decrease upon entering the $N_{tb}$ phase and $<P_4(\cos\theta)>$ becomes negative providing a direct confirmation of the conical molecular orientational distribution. The heliconical tilt angle, estimated from the orientational distribution functions (ODFs), in all cases increases from zero at the N - $N_{tb}$ transition to approximately 27° at about 40 K below the transition, in excellent agreement with freeze fracture transmission electron microscopy results of Chen and the birefringence results of Meyer. The growth of the tilt angle in the $N_{tb}$ phase follows a single power law with exponents between ~0.09 ± 0.01 to -0.12 ± 0.01, which is far from the expected tricritical or mean field exponents of 0.25 or 0.5. The temperature dependence of the tilt angle calculated from the ODFs also is in good qualitative agreement with the values estimated from optical studies of their rope-like textures within adjacent blocks of left- and right-handed twist in homogeneously aligned cells.

**Significance**: In a nematic liquid crystal (LC), rod-like organic molecules are orientationally ordered along a direction that becomes its optic axis. This type of LC was discovered more than 125 years ago by Reinitzer and is currently used in practically all LC flat panel displays and electrooptical devices. A new type of nematic LC, the *twist-bend nematic,* was recently discovered. It has the potential of leading to the next generation of, electrooptical devices that are expected to be about one hundred times faster than current devices. The structure of this phase is heliconical, in which the optic axis surprisingly adopts a helical trajectory even though the constituent molecules are achiral. The twist-bend phase is only the fifth nematic phase to be discovered and presents very interesting and unexpected scientific phenomena for researchers.

## 1. Introduction:

The *nematic* (N) phase is the most widely studied liquid crystal phase and is used in the ubiquitous flat panel displays and other electrooptical applications. This phase is characterized[1] by long-range orientation order of the molecules' symmetry axis along the *director* (a pseudo-vector) **n**, as shown in Fig. 1(a). When the molecules are chiral (i.e., lack mirror symmetry), the *cholesteric* phase is obtained. Here, the local **n** describes a helical trajectory, Fig. 1(b), about a direction perpendicular to itself. Recently, a very different kind of chiral nematic phase was discovered in systems of calamitic bent-shape bimesogens. It differs from the previously known cholesteric phase as the local director **n** precesses about the helical axis **N** at an oblique angle α [Fig. 1(c)] and lies on the surface of a cone. This heliconical phase is called the *twist-bend nematic* ($N_{tb}$) phase and is known to be a consequence of a negative bend elastic constant. Although the $N_{tb}$ phase was originally predicted on the basis of theoretical considerations[2] and simulations[3] for bent-core (or, banana-shape) molecules, it was first discovered in calamitic bimesogens[4,5,6] and then in two systems of bent-core (banana-shaped) molecules by the Clark[7] and Kumar[8] groups.

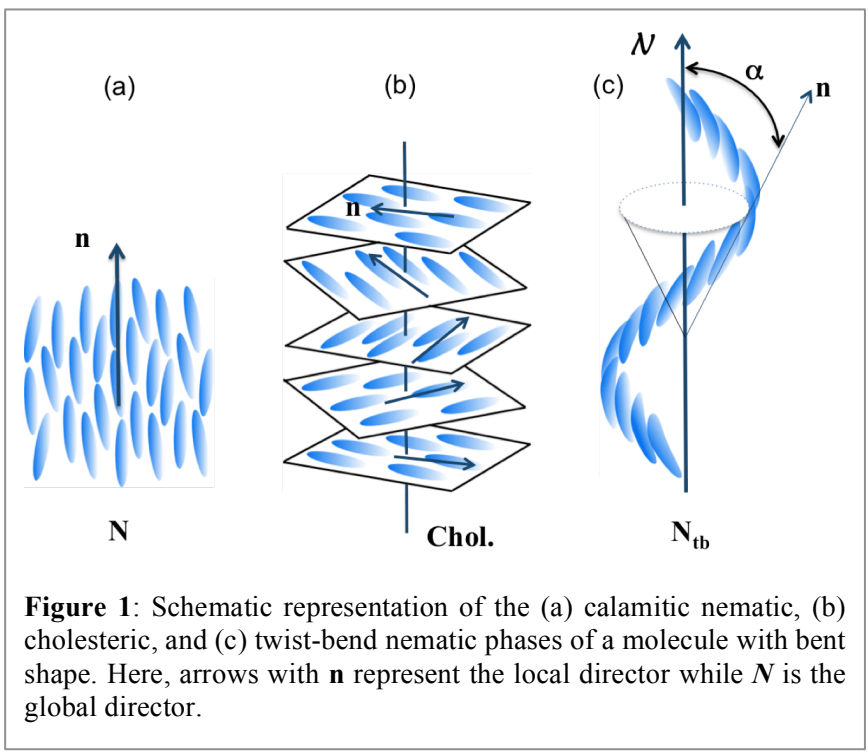

**Figure 1**: Schematic representation of the (a) calamitic nematic, (b) cholesteric, and (c) twist-bend nematic phases of a molecule with bent shape. Here, arrows with **n** represent the local director while **N** is the global director.

Using a simple phenomenological model, Dozov[2] found that systems of rigid achiral bent-core molecules are conducive to the natural formation of local bend distortions of the director at a relatively low cost in energy. He predicted a new nematic phase possessing either an *oscillating splay-bend* or a *heliconical twist-bend* ($N_{tb}$) director configuration at temperatures below the N phase. A negative value of the bend elastic constant was a prerequisite for the formation of these phases. Interestingly, computer simulations of Memmer, *et al.*,[3] which used Gay-Berne potential between molecules of $C_{2v}$ symmetry, also predicted this phase with no smectic-like layer structure at intermediate temperatures. Shamid, *et al.*,[9] incorporated polar order in their model and found that, under proper conditions, such molecules may form the nematic, cholesteric, biaxial nematic, and a polar nematic phase with transverse polar order. The latter could either be a splay-bend or a twist-bend phase predicted by Dozov.

All known calamitic dimers exhibiting the $N_{tb}$ phase consist of two mesogenic units linked by a flexible spacer consisting of an odd number of carbons that give the molecule a bent shape. For example, two cyanobiphenyl (CB) moieties linked by alkyl chains of three, five, or seven carbon atoms (*e.g.*, CB7CB, two CB moieties connected by seven methylenes) exhibit the N-$N_{tb}$ transition, while the same set of moieties linked with even number methylenes exhibit no such transition. It should be noted that initially, the new nematic phase was called as the $N_x$ phase. Cestari, *et al.*,[6] performed an extensive study of CB7CB and revealed that, in spite of exhibiting smectic like defects, the $N_{tb}$ phase was devoid of any smectic order. Their data suggested that the N-$N_{tb}$ phase transition was near a tricritical point due to a coupling between the nematic and heliconical tilt orders. Experiments showed that the heat capacity critical exponent was 0.5 consistent with tricritical behavior. The tricritical nature was further confirmed[10] by adiabatic calorimetric studies of the mixtures of the bimesogens (CB7CB) and a monomer nematic 4-pentyl-4'-cyanobiphenyl (CB5).

Two freeze fracture transmission electron microscopy (FFTEM) studies[11,12] of two compounds have revealed a periodicity of ~ 8-9 *nm* in the $N_{tb}$ phase of calamitic bimesogens. This so far remains to be the only direct evidence of a periodic structure that is consistent with the observed smectic-like defects. While this length scale is accessible in small-angle x-ray scattering experiments, no such evidence has been found in spite of several very diligent efforts[12,13]. Furthermore, the FFTEM textures also reveal other periodicities[11] as small as 3.4 *nm*, *i.e.,* approximately equal to the molecular length, which are not trivial to understand.

In order to gain further insight into the molecular organization in the $N_{tb}$ phase and the nature of the N-$N_{tb}$ transition, we investigated several mixtures of CB6CB and CB7CB at CB6CB concentrations below 25 wt%. The results of polarized optical microscopy (POM) and x-ray diffraction (XRD) measurements of the evolution of structure and the orientational order parameters $<P_2(\cos\theta)>$ and $<P_4(\cos\theta)>$ across the N-$N_{tb}$ transition are presented here. These order parameters reveals a conical (or, volcano-like[14]) molecular orientational distribution (ODF) function in the $N_{tb}$ phase and allow a direct estimation of its cone angle, an order parameter of the $N_{tb}$ phase. The values and thermal evolution of the optically measured tilt of the local director with respect to the helical axis (*i.e.*, the cone angle) are in good agreement with our x-ray results as well as those previously reported[15] values from optical birefringence. Furthermore, the temperature dependence of the helical pitch is in conformity with theoretical expectations of Dozov.

## 3. Results and Discussions:

The N phase in different mixtures was confirmed through its characteristic POM schlieren textures, the presence of two- and four-brush disclinations, and the director fluctuations. The $N_{tb}$ phase was similarly characterized by its broken focal conic and rope like textures[5]. The phase diagram for CB7CB + CB6CB system is presented in Fig. 2. Schlieren textures of the N phase and broken fan and rope like textures of the $N_{tb}$ phase of pure CB6CB and CB7CB are shown in Fig. 3. Similar textures were observed for the 5, 10, and 15 wt% CB6CB samples. The $N_{tb}$ phase is observed up to 20 wt% CB6CB. Aligned greyish textures appeared in the N schlieren textures of the samples with ≥ 20 wt% CB6CB concentration, and more readily at higher concentrations. This appears to be a consequence of phase separation into CB6CB-rich and CB7CB-rich regions mainly due of conformational immiscibility of the two types of bimesogenic molecules, which limited our study to < 20 wt% of CB6CB.

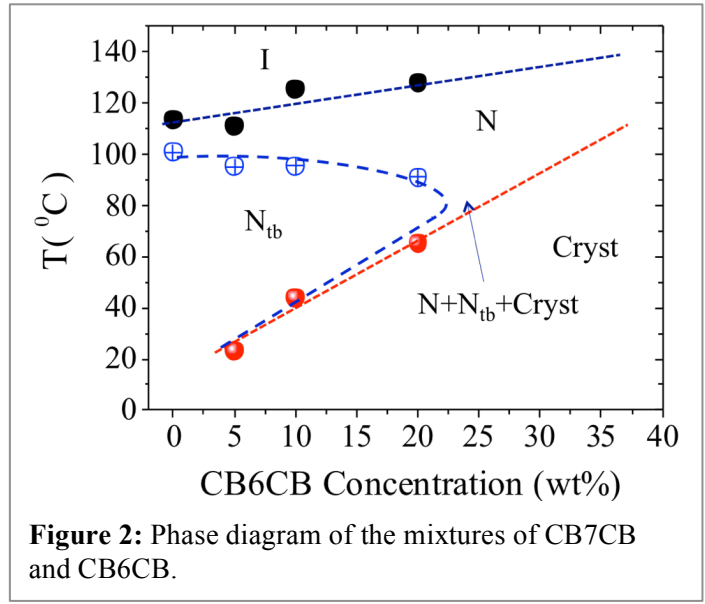

**Figure 2:** Phase diagram of the mixtures of CB7CB and CB6CB.

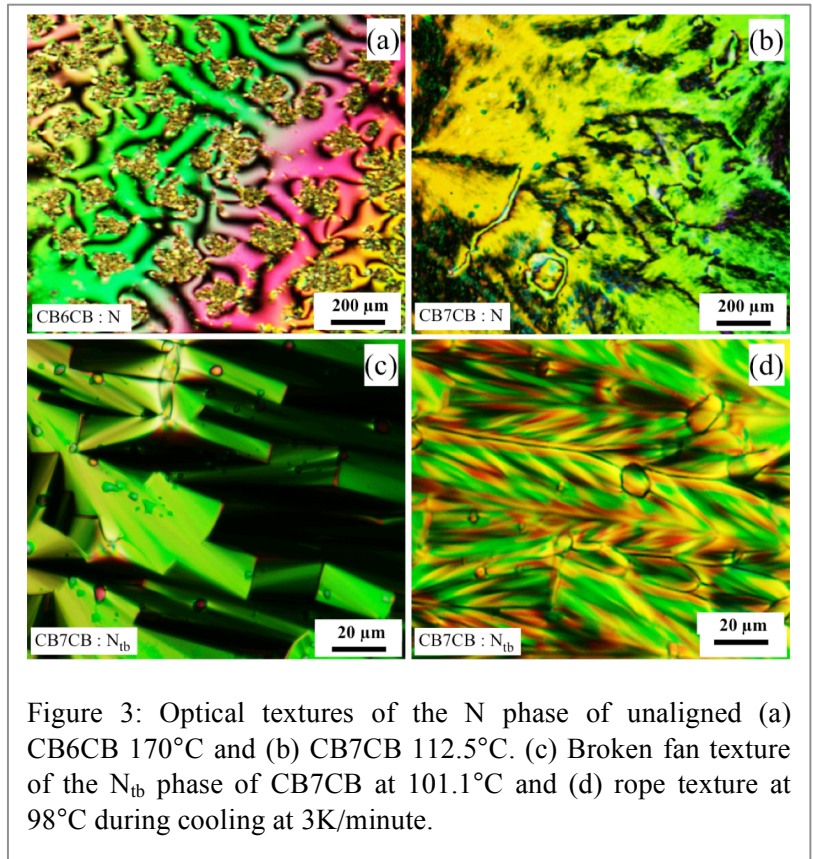

Figure 3: Optical textures of the N phase of unaligned (a) CB6CB 170°C and (b) CB7CB 112.5°C. (c) Broken fan texture of the $N_{tb}$ phase of CB7CB at 101.1°C and (d) rope texture at 98°C during cooling at 3K/minute.

Multi-strand rope-like textures, with apparent helical winding separated by dark lines into locally parallel blocks [Figs. 3(d) and 4(a)], readily formed in homogeneously aligned cells of all samples in the $N_{tb}$ region. The quality of alternating right- and left-handed strands in adjacent blocks varied from sample to sample, from one region of a cell to another, and with temperature.

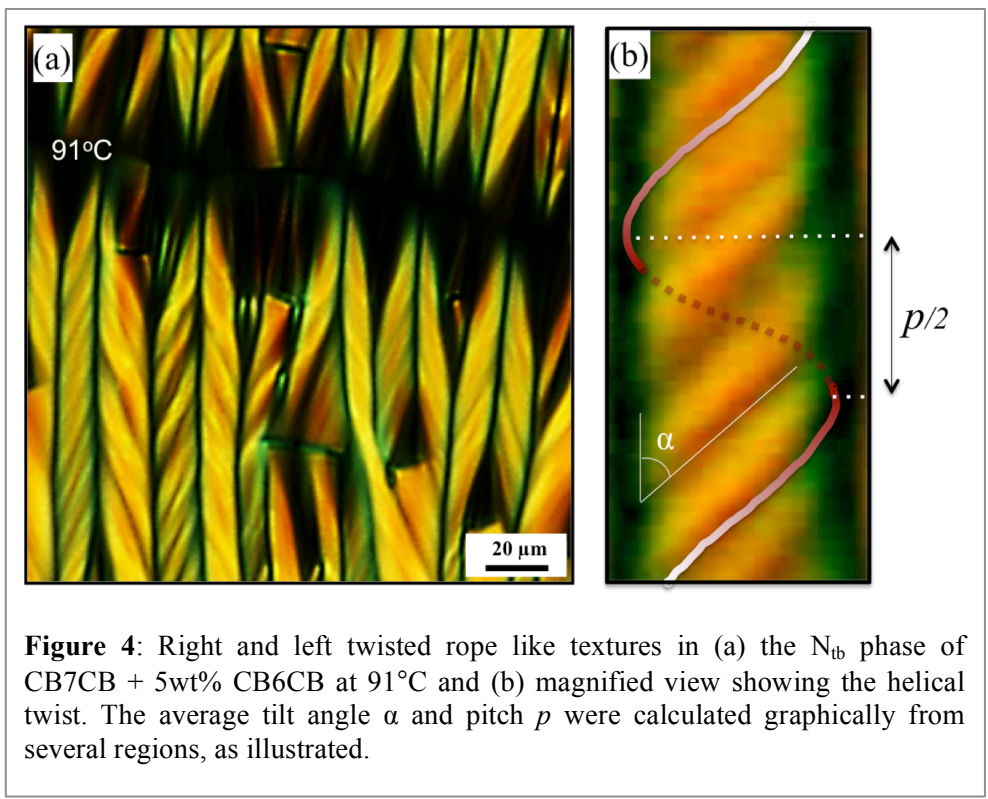

**Figure 4**: Right and left twisted rope like textures in (a) the $N_{tb}$ phase of CB7CB + 5wt% CB6CB at 91°C and (b) magnified view showing the helical twist. The average tilt angle α and pitch $p$ were calculated graphically from several regions, as illustrated.

High-quality rope-like textures emerged parallel to the rubbing direction in homogeneous cells, at ~ 1 - 2 K below the transition to the $N_{tb}$ phase. A representative doubly degenerate helical (rope-like) texture in the $N_{tb}$ phase of the 5 wt% CB6CB sample is shown in Fig. 4. It was often possible to find several regions of high-quality texture, where the average optical pitch/periodicity $p$ parallel to the rope axis (i.e., the dark lines) and the (tilt) angle α between the strands and the axis of helix could be determined as functions of temperature.

The quality of the textures often limited our ability to accurately measure the tilt angle and the pitch. The rope/block widths remained essentially temperature-independent and comparable to the cell thickness. The blocks are believed[11,16] to arise from Helfrich-Hurault undulation instability[17,18] previously observed in the smectic and cholesteric phases. However, the internal structure of the blocks, i.e. the pitch of the strands and their tilt, do not depend on cell parameters and are temperature dependent, and therefore appear to reflect an intrinsic property of the phase. The pitch $p$ and the tilt angle $\alpha$ measured graphically from POM textures of CB7CB are plotted in Fig. 5. The pitch increases from ~10 to 40 μm showing the unwinding of helix and the tilt α decreased rapidly from ~ 40 to 22° as the N phase was approached from below. The temperature dependence of α for CB7CB is in excellent agreement with the values indirectly calculated[15] from the birefringence of the $N_{tb}$ phase, confirming that our measurements reflect an intrinsic property of the $N_{tb}$ phase. The agreement leads one to speculate that the nanoscale periodic structure manifests itself, in yet unknown way, into microscale textural features that depend and reflect the actual value of the tilt angle.

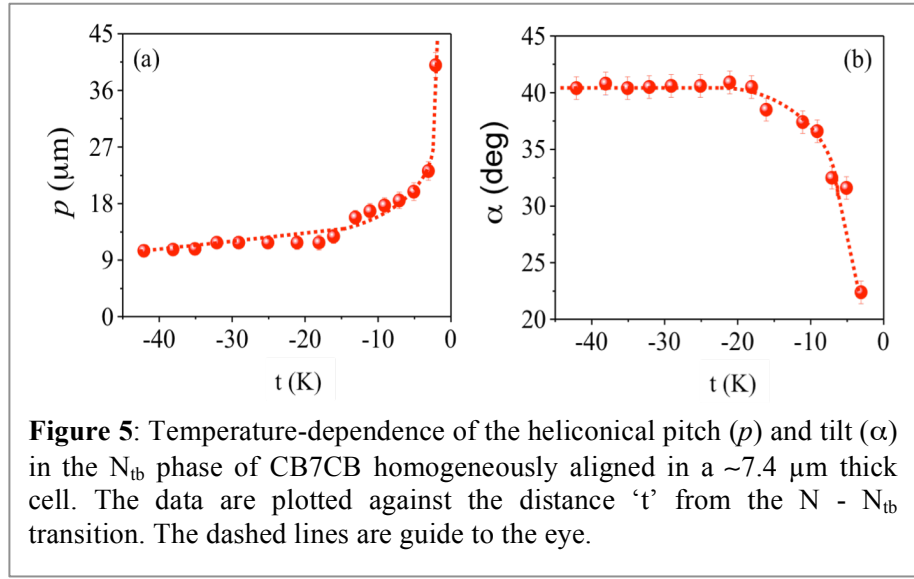

**Figure 5**: Temperature-dependence of the heliconical pitch ($p$) and tilt ($\alpha$) in the $N_{tb}$ phase of CB7CB homogeneously aligned in a ~7.4 μm thick cell. The data are plotted against the distance 't' from the N - $N_{tb}$ transition. The dashed lines are guide to the eye.

The x-ray diffraction patterns consist of two pairs of orthogonal diffuse crescents, one at small angles and the other in the wide-angle region, confirming an aligned N phase, Fig. 6. The length scale (or, *d*-spacing) corresponding to these reflections of CB7CB were ~ 12 Å and ~ 4.45 Å in agreement with previously reported values[6]. The temperature dependence of the effective length scale obtained from the small angle peak shows a small discontinuity at the N-$N_{tb}$ transition consistent with its weakly first order nature; the lateral length scale shows a continuous evolution. The value 12Å was about half of the bimesogen's length (26Å) suggesting an intercalated organization of molecules as depicted in Fig. 7. Such intercalation where the aromatic part of one molecule is adjacent to the aliphatic segment of the next molecule, would give rise to string-like association of mesogens with much reduced electron density contrast along their length, thus reducing the intensity of the small angle peak. The intensity of the small-angle reflection was indeed significantly lower than the wide-angle peak in comparison to the regular N phase of most other compounds where the small angle peak is stronger.

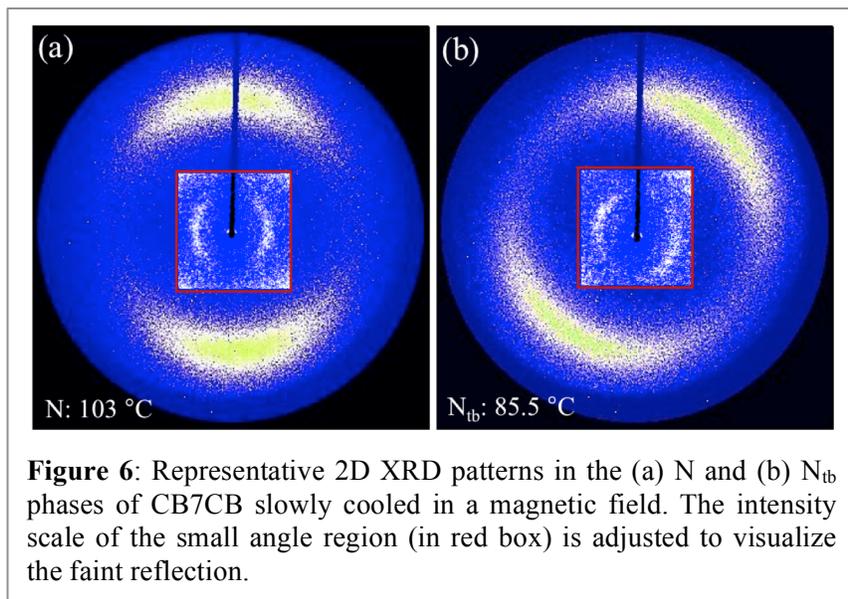

**Figure 6**: Representative 2D XRD patterns in the (a) N and (b) $N_{tb}$ phases of CB7CB slowly cooled in a magnetic field. The intensity scale of the small angle region (in red box) is adjusted to visualize the faint reflection.

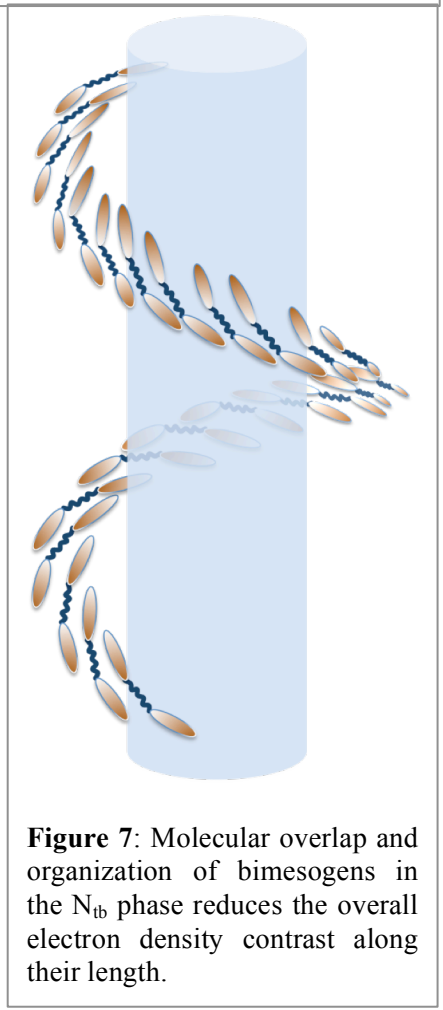

**Figure 7**: Molecular overlap and organization of bimesogens in the $N_{tb}$ phase reduces the overall electron density contrast along their length.

The diffraction pattern always rotated unpredictably after the transition to the $N_{tb}$ phase, Fig. 6(b), relative to the pattern in the N phase, as in previously published x-ray patterns[4,13,19]. It is very likely caused by two factors: (i) the formation of the heliconical organization of the mesogens which disrupts the director orientation in the N phase, and (ii) the reduced ability of the magnetic field to maintain the director orientation parallel to itself as a consequence of helical structure and reduced diamagnetic anisotropy (accompanying reduced order) of the $N_{tb}$ phase. Clearly, the lack of sharp small angle peak confirms an absence of smectic-like layering which was expected based on the appearance of focal conic textures in this phase.

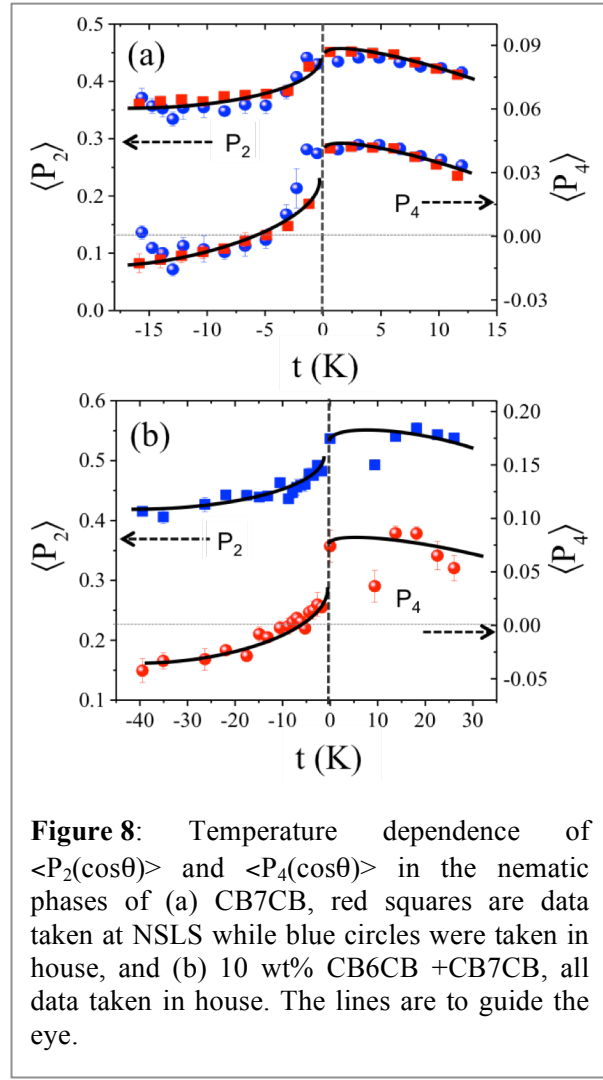

**Figure 8**: Temperature dependence of $<P_2(\cos\theta)>$ and $<P_4(\cos\theta)>$ in the nematic phases of (a) CB7CB, red squares are data taken at NSLS while blue circles were taken in house, and (b) 10 wt% CB6CB +CB7CB, all data taken in house. The lines are to guide the eye.

The line shape of the wide-angle reflections was analyzed to calculate[20] the nematic orientational order parameters $<P_2(\cos\theta)>$, $<P_4(\cos\theta)>$, and $<P_6(\cos\theta)>$ using the numerical inversion method of Davidson, *et al.*[21]. The parameter $<P_6>$ was nearly zero in all cases. In the case of CB6CB, order parameters $<P_2>$ and $<P_4>$ increased with decreasing temperature in the N phase, and $<P_2>$ was close to typical value of ~ 0.6 for the N phase. The temperature-dependence of the two order parameters for CB7CB samples calculated from, both, the data taken in house (blue circles) and at NSLS (red squares) are shown in Fig. 8(a). While the two sets of results are in very good agreement, the NSLS data apparently is of higher quality. In the N phase of CB7CB, both <$P_2$> and <$P_4$> are positive and increase from ~ 0.40 to 0.45 and ~ 0.03 to 0.04, respectively, as the temperature is lowered. The values of <$P_2$> in the $N_{tb}$ phase are in agreement with those obtained[22] from the $^{13}$C 2D NMR technique. We are unable to compare the values of <$P_4$> with results from other experiments, as none exist at this time. After a significant initial drop in their values at the N – $N_{tb}$ transition, both order parameters continued to decrease. While <$P_2$> remained above + 0.36, the values of <$P_4$> became negative soon after entering the $N_{tb}$ phase, approaching ~ − 0.02 at ~ 15 K below the transition. The 5 and 10 wt% samples show the same trends as pure CB7CB in the N and $N_{tb}$ phases with somewhat different numerical values. Fig. 8(b) presents the two order parameters for the 10 wt % sample. Most notably, the value of <$P_4$> becomes negative a few degrees after entering the $N_{tb}$ phase. The negative value of <$P_4$> has importance as it confirms the conical distribution of the local director in the $N_{tb}$ phase, as discussed below.

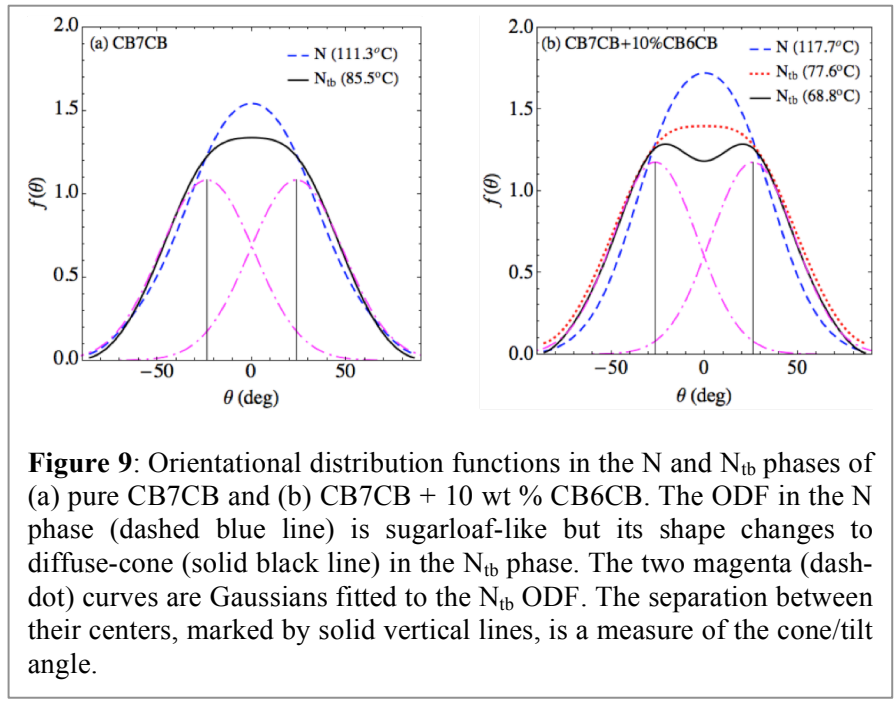

**Figure 9**: Orientational distribution functions in the N and $N_{tb}$ phases of (a) pure CB7CB and (b) CB7CB + 10 wt % CB6CB. The ODF in the N phase (dashed blue line) is sugarloaf-like but its shape changes to diffuse-cone (solid black line) in the $N_{tb}$ phase. The two magenta (dash-dot) curves are Gaussians fitted to the $N_{tb}$ ODF. The separation between their centers, marked by solid vertical lines, is a measure of the cone/tilt angle.

The ODFs in the two nematic phases of CB7CB and the 10 %CB6CB sample, shown in Fig. 9, were calculated from Legendre functions <$P_2$> and <$P_4$> obtained from x-ray experiments.

The ODFs in the N phase, where both order parameters are positive, are sugarloaf like shown by dashed blue curve. But it has a conical (or, volcano-like) shape in the $N_{tb}$ phase providing the first direct evidence of conical distribution of the local director as previously predicated[23]. These ODF plots represent a cross-section of the distribution in a plane containing the axis of the cone (i.e., the helix), and the two high points on the ODF depend on the tilt angle α of the local director with respect to the global director $N$, or the cone axis. We estimated α by fitting a sum of two Gaussians shown by dot-dash curves in Fig. 9, to the ODF at different temperatures. The angle α is one half of the separation between the centers of the Gaussians, which is plotted in Fig. 10 against temperature for three samples. The tilt angle increases from zero at the N to $N_{tb}$ transition and attains a value between 26.5° and 28° for the samples studied. The temperature dependence of the tilt angle fits well to a single power law with the exponent ranging from 0.10 ± 0.02 for the 10 wt% sample to 0.12 ± 0.02 for the 5wt % sample. Since the cone angle is one of the order parameters of the $N_{tb}$ phase, its exponent is expected to be either 0.25 for tricritical or 0.5 for the mean-field behavior. Our values are not in agreement with these expectations. This is not surprising because the transition in these materials has been shown to be weakly first order and the region close to the transition was not explored in these measurements.

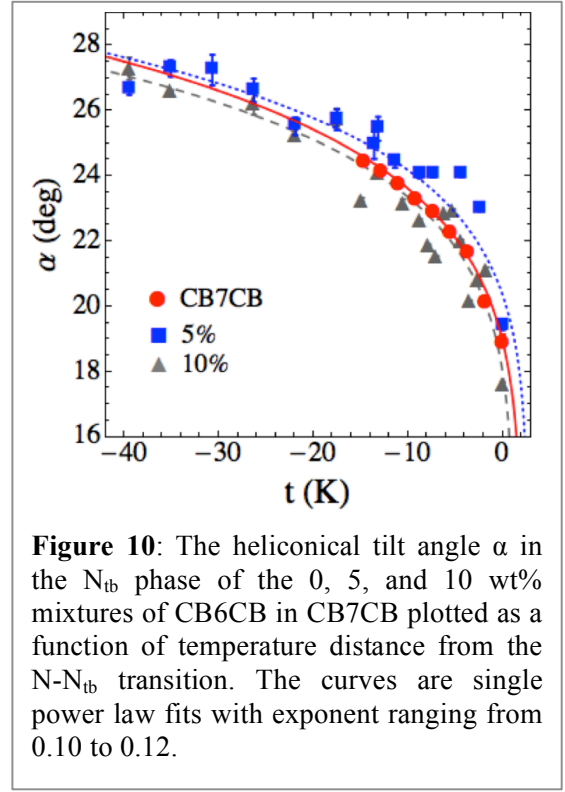

**Figure 10**: The heliconical tilt angle α in the $N_{tb}$ phase of the 0, 5, and 10 wt% mixtures of CB6CB in CB7CB plotted as a function of temperature distance from the N-$N_{tb}$ transition. The curves are single power law fits with exponent ranging from 0.10 to 0.12.

To conclude, the temperature dependence of the order parameters $<P_2>$ and $<P_4>$ has been measured in the N and $N_{tb}$ phases of CB7CB and its mixtures with CB6CB. A negative value of $<P_4>$ confirms a conical local director configuration in the $N_{tb}$ phase. The ODF has a sugarloaf shape in the N phase that becomes volcano-like in the $N_{tb}$ phase. Thermal evolution of the cone angle, directly measured via the ODFs, is in good qualitative agreement with the values estimated from rope-like optical textures, and previously reported values. There is a puzzle as to how the cone angle measured at small length scales via x-ray diffraction and at several orders of magnitude larger optical scales behave the same and the physics that links these two together.

## 2. Experimental Methods:

Binary mixtures of symmetric liquid crystal dimers[24] CB6CB and CB7CB were prepared by dissolving pre-weighed amounts in chloroform, allowing the solvent to slowly evaporate at room temperature, and then keeping the sample under vacuum for 24 hours. A polarizing optical microscope (Olympus BX51) equipped with a CCD camera (SPOT, Diagnostic Instruments, Inc.) was used to record optical textures during heating and cooling cycles. Glass substrates were coated with polyimide SE7492 films, mechanically rubbed, and assembled in an anti-parallel manner to prepare homogenously aligned cells. Unaligned and homogeneously aligned samples of 4.0 - 7.5 μm thickness were placed in a Mettler Hotstage (FP82HT) with thermal stability of ± 0.1C°. Cells with homogeneous alignment produced good quality rope-like textures that were used to estimate the optical pitch and the tilt angle of the heliconical structure at different concentrations and temperatures.

For x-ray diffraction, samples were filled in 1.5 mm diameter quartz capillaries and flame sealed. For liquid crystal alignment and temperature control, the filled capillaries were placed in an in-situ magnetic field of approximately 2.5 kG produced by a pair of rare-earth permanent magnets mounted inside an INSTEC HS402 hot stage. X-ray experiments were performed using a microfocus Rigaku Screen Machine with Copper anode ($\lambda$ = 1.542 Å) and a Mercury 3 CCD detector positioned at ~ 64 *mm* from the sample, and at the beamline X27C of the National Synchrotron Light Source (NSLS) using 1.371 Å x-rays where the XRD patterns were acquired with a MAR CCD detector (resolution 160 × 160 μm$^2$) placed at ~222.4 *mm* from the sample. The data were calibrated against silver behenate and silicon standards traceable to the National Institute of Standards and Technology. Background scattering was recorded with an empty capillary in the sample position and subtracted from actual scattering data. The corrected data were analyzed using FIT2D software[25].

## 5. Acknowledgement:

This work was supported by the Basic Energy Science program of the Office of Science, Department of Energy award SC-0001412. They synthesis work of J.-K. S. was supported by the National Research Foundation of Korea (NRF) grant funded by MSIP (No. 2014R1A2A1A11054392). We thank N. Clark, A. Rey, R. Kamien, M. Copic, J. Selinger, and I. Dozov for very helpful discussions.